\documentclass[hyphens]{article}

\PassOptionsToPackage{numbers,compress}{natbib}

\usepackage[preprint]{neurips_2025}

\makeatletter
\renewcommand{\@noticestring}{}
\makeatother

\usepackage[utf8]{inputenc}
\usepackage[T1]{fontenc}

\usepackage[hyphens]{url}
\usepackage{hyperref}

\usepackage{breakurl}
\usepackage{booktabs}
\usepackage{amsfonts,amsmath}
\usepackage{nicefrac}
\usepackage{microtype}
\usepackage{multirow}
\usepackage{makecell}
\usepackage{graphicx}
\usepackage{subcaption}
\usepackage{todonotes}
\usepackage{placeins}
\usepackage{CJKutf8}
\usepackage{booktabs}
\usepackage{colortbl}
\usepackage{framed}
\usepackage{listings}
\usepackage{xcolor}
\usepackage{float} 
\usepackage{relsize}
\usepackage{listings}
\usepackage{xcolor}
\lstdefinestyle{sqlstyle}{
  language=SQL,
  basicstyle=\footnotesize\ttfamily,
  keywordstyle=\color{blue}\bfseries,
  commentstyle=\color{gray},
  stringstyle=\color{red},
  showstringspaces=false,
  breaklines=true,
  frame=single,
  columns=fullflexible
}

\definecolor{lavender}{RGB}{204,204,255}

\makeatletter
\renewcommand\@biblabel[1]{\hfill[#1]}
\makeatother

\title{M3: Conversational LLMs Simplify Secure Clinical Data
Access, Understanding, and Analysis}

\author{%
  Rafi Al Attrach$^{1,2,*}$,
  Pedro Moreira$^{1,3,*}$,
  Rajna Fani$^{1,2,*}$,\\
  \textbf{Renato Umeton$^{1,4}$,
  Amelia Fiske$^{2}$,
  Leo Anthony Celi$^{1,5,6,\dagger}$}\\
  \\
  $^1$Massachusetts Institute of Technology, $^2$Technical University of Munich,\\
  $^3$Universitat Pompeu Fabra, $^4$St. Jude Children's Research Hospital,\\
  $^5$Beth Israel Deaconess Medical Center,\\
  $^6$Harvard T.H. Chan School of Public Health, Department of Biostatistics\\
  \\
  $^*$Co-first authors. $^\dagger$Corresponding author: \texttt{lceli@mit.edu}.\\
  \small Accepted at the Machine Learning for Health (ML4H) Symposium 2025.
}
\def\github{\raisebox{-1.5pt}{\includegraphics[height=1.05em]{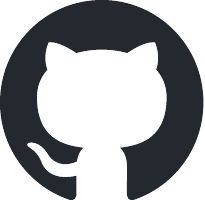}}}
\newcommand{\ghlink}{https://github.com/rafiattrach/m3}

\def\pypi{\raisebox{-1.5pt}{\includegraphics[height=1.05em]{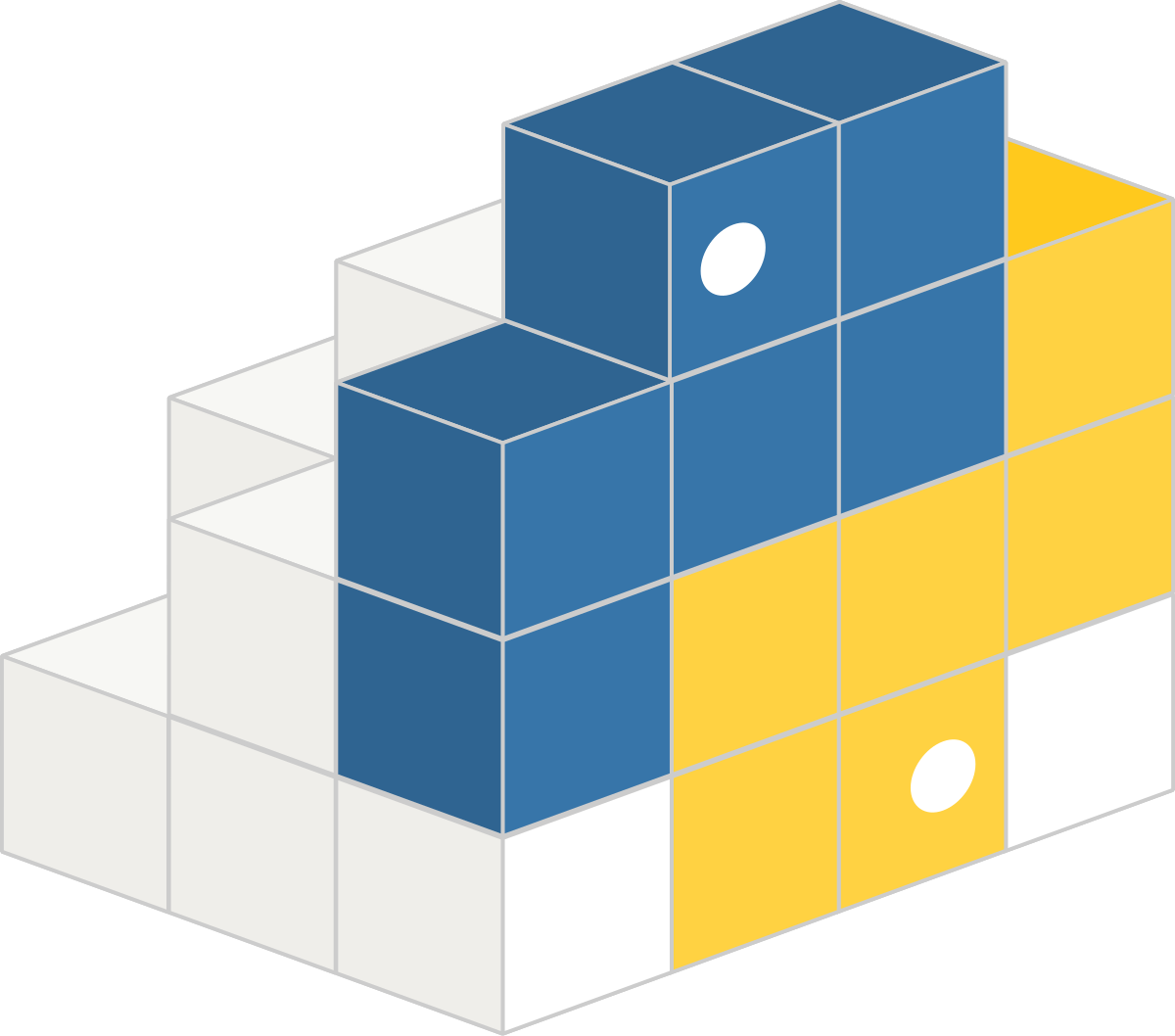}}}
\newcommand{\pypilink}{https://pypi.org/project/m3-mcp}

\def\website{\raisebox{-1.5pt}{\includegraphics[height=1.05em]{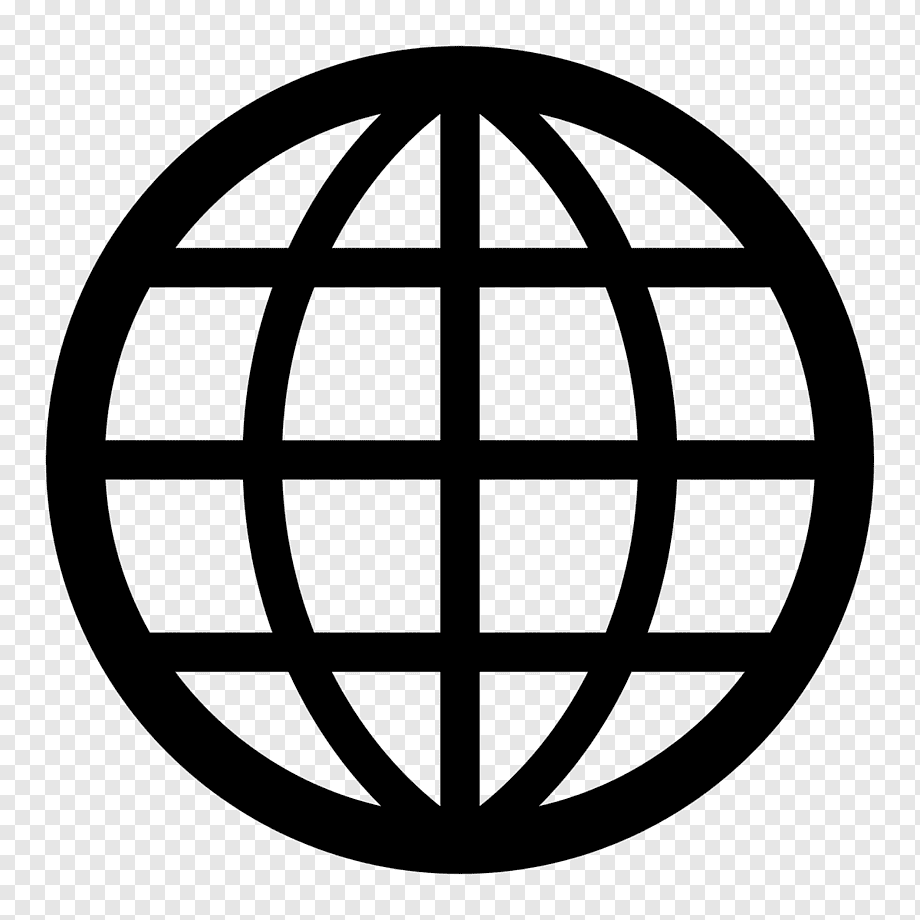}}}
\newcommand{\websitelink}{https://rafiattrach.github.io/m3/}

\begin{document}
\maketitle
\begin{center}
\vspace{-1cm}
\begin{tabular}{rl}
\github & \url{\ghlink}\\
\pypi & \url{\pypilink}\\
\website & \url{\websitelink}\\
\end{tabular}
\end{center}
\begin{abstract} 

Large-scale clinical databases offer opportunities for medical research, but their complexity creates barriers to effective use. The \textbf{Medical Information Mart for Intensive Care (MIMIC-IV)}, one of the world's largest open-source electronic health record databases, traditionally requires both SQL proficiency and clinical domain expertise. We introduce \textbf{M3}, a system that enables natural language querying of MIMIC-IV data through the \textbf{Model Context Protocol}. With a single command, M3 retrieves MIMIC-IV from PhysioNet, launches a local SQLite instance or connects to hosted BigQuery, and allows researchers to pose clinical questions in plain English. We evaluated M3 using samples from the EHRSQL 2024 benchmark with two language models. On one hundred answerable questions, the proprietary Claude Sonnet 4 achieved 94\% accuracy and the open-weights gpt-oss-20B (deployable locally on consumer hardware) achieved 93\%; on a matched sample of one hundred unanswerable questions, where correct behavior is to abstain rather than produce SQL, gpt-oss-20B correctly abstained on 69\%. Both models translate natural language into SQL, execute queries against MIMIC-IV, and return structured results alongside the underlying query for verification. Error analysis revealed that most failures stemmed from complex temporal reasoning or ambiguous question phrasing rather than fundamental architectural limitations. The comparable performance of a smaller open-weights model demonstrates that privacy-preserving local deployment is viable for sensitive clinical data analysis. M3 lowers technical barriers to critical care data analysis and is designed with security measures including OAuth2 authentication, query validation, and audit logging.
\end{abstract}

\section{Introduction}
\subsection{The Challenge of Analyzing Large-Scale Clinical Databases}

The digital transformation of healthcare has generated vast quantities of electronic health record (EHR) data, creating resources for medical research that offer insights into disease patterns, treatment efficacy, and patient outcomes \cite{evans2016electronic}. However, barriers to using these datasets remain high due to inherent data complexity and required technical skills. Clinical databases typically consist of numerous interconnected tables with specific definitions and coding schemes. Navigating and extracting meaningful information requires proficiency in Structured Query Language (SQL), thorough understanding of database schema, and knowledge of how data points connect temporally and semantically.

This technical requirement limits the pool of researchers able to leverage these resources, potentially impeding innovation in clinical process improvement. An interdisciplinary gap exists where clinical experts who formulate critical research questions may be disconnected from the data extraction process, which often falls to data scientists or programmers. Tools that bridge this divide by simplifying data access have become important, with some already in use in academic medical centers and clinical settings \cite{statnewsSvelte,jiang2023health,doi:10.1056/AIcs2300191}.

Anthropic's Model Context Protocol (MCP) \cite{anthropic2024mcp} provides a standardized framework for managing AI model interactions with external software tools and data sources, offering an approach to address these accessibility challenges through secure and controlled interfaces.

\subsection{The Role of MIMIC-IV in Critical Care Research}

MIMIC-IV stands as a cornerstone publicly available database for critical care research. Developed by the MIT Laboratory for Computational Physiology, this dataset contains de-identified health data associated with patients admitted to intensive care units or the emergency department at Beth Israel Deaconess Medical Center. The current release is MIMIC-IV v3.1 \cite{mimicivphysionet}; as reported by PhysioNet, as of v3.0 the dataset contains approximately 364,627 unique individuals, 546,028 hospitalizations, and 94,458 ICU stays. The dataset includes patient demographics, vital signs, laboratory results, medications, procedures, and more.

MIMIC-IV is utilized in the research community for developing clinical prediction models, understanding disease trajectories, evaluating treatment interventions, and improving patient care in critical settings. The availability of MIMIC-IV through PhysioNet \cite{goldberger2000physionet}, which provides access through Google BigQuery for the full dataset, enhances research transparency and reproducibility. The public, albeit credentialed, nature of MIMIC-IV enables research groups to work with standardized, high-fidelity clinical data, fostering collaboration and building upon prior work.

\subsection{Introducing M3: Objectives and Contributions}

This paper introduces M3, a project developed to address challenges of accessing and analyzing MIMIC-IV data. The primary objective is to transform how researchers interact with this medical data resource by enabling natural language querying facilitated by AI assistance.

The key contributions are: (1) A software framework specifically designed to simplify data access for the MIMIC-IV database. (2) A concrete, auditable integration of an LLM tool layer with MIMIC-IV using the Model Context Protocol, providing standardized tool exposure, execution logging, and access-control boundaries. We frame this as an engineering integration of existing technology rather than a novel protocol contribution. (3) Empirical evaluation on the MIMIC-IV demo (SQLite) with two language models, demonstrating feasibility of natural language-to-SQL translation in a clinical context; the full-scale BigQuery backend is supported architecturally, with empirical benchmarking on the full dataset identified as future work. (4) A step toward lowering the technical barrier to entry for MIMIC-IV research, making the data more accessible to a broader range of researchers.

M3 represents a concrete application of Natural Language Interface and text-to-SQL research, tailored to a high-impact medical dataset.

\begin{figure}[H]
    \centering
    \includegraphics[width=0.7\linewidth]{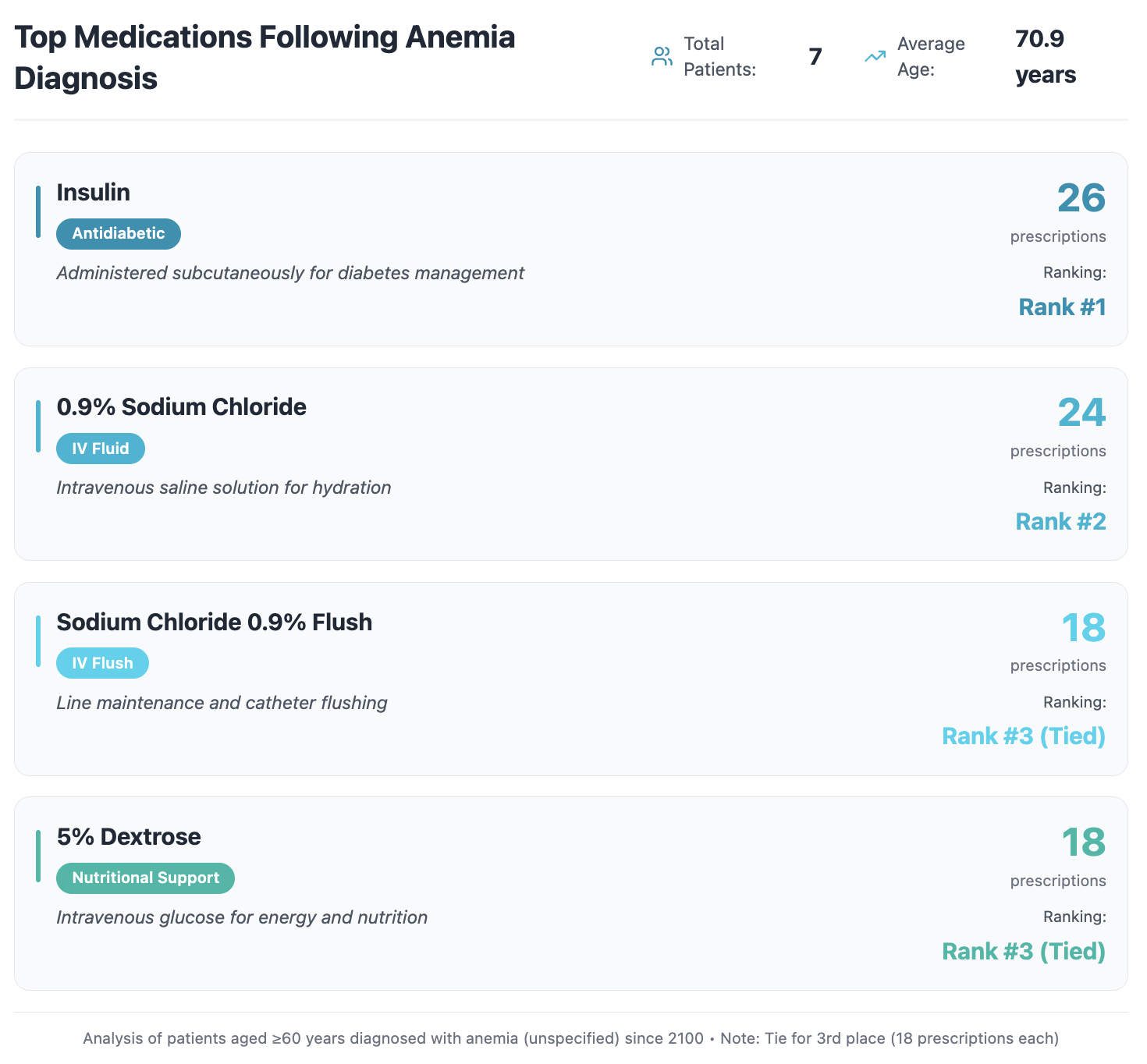}
    \caption{\textbf{Results of a complex natural language query.}
    Query: \textit{``Among patients who were diagnosed with anemia, unspecified since 2100, what are the top three most commonly prescribed medications that followed during the same hospital visit for patients in their 60 or above?''} The M3 system translated this natural language question into appropriate SQL and returned structured results.}
    \label{fig:complex_query}
\end{figure}

We evaluate M3 using samples from the EHRSQL 2024 test set \cite{ehrsql2024github}, a benchmark designed for assessing natural language-to-SQL performance in clinical contexts based on the publicly available MIMIC-IV demo (v2.2) \cite{johnson2023mimiciv_demo}. Figure~\ref{fig:complex_query} shows the result of a complex query involving multiple temporal and clinical constraints, obtained through the M3 system powered by Claude Sonnet 4 \cite{anthropic2025claude4} via MCP.

For comparison, Listing~\ref{lst:correct-sql} shows the corresponding SQL query that a researcher would otherwise need to construct manually.

\begin{lstlisting}[style=sqlstyle, caption={Correct SQL query for the complex natural language question shown in Figure~\ref{fig:complex_query}. This query requires knowledge of table relationships, temporal logic, and SQL windowing functions that M3 generates automatically from the natural language input. The query targets the EHRSQL 2024 benchmark SQLite schema \cite{ehrsql_github}, in which the \texttt{admissions} table exposes a direct \texttt{age} column; this is a benchmark-specific transformation of the underlying MIMIC-IV data.}, label={lst:correct-sql}]
SELECT T3.drug
FROM (
  SELECT T2.drug, DENSE_RANK() OVER (ORDER BY COUNT(*) DESC) AS C1
  FROM (
    SELECT admissions.subject_id, diagnoses_icd.charttime, admissions.hadm_id
    FROM diagnoses_icd
    JOIN admissions ON diagnoses_icd.hadm_id = admissions.hadm_id
    WHERE diagnoses_icd.icd_code = (
      SELECT d_icd_diagnoses.icd_code
      FROM d_icd_diagnoses
      WHERE d_icd_diagnoses.long_title = 'anemia, unspecified'
    )
    AND strftime('%Y', diagnoses_icd.charttime) >= '2100'
  ) AS T1
  JOIN (
    SELECT admissions.subject_id, prescriptions.drug, prescriptions.starttime, admissions.hadm_id
    FROM prescriptions
    JOIN admissions ON prescriptions.hadm_id = admissions.hadm_id
    WHERE admissions.age >= 60
    AND strftime('%Y', prescriptions.starttime) >= '2100'
  ) AS T2
  ON T1.subject_id = T2.subject_id
  WHERE T1.charttime < T2.starttime
    AND T1.hadm_id = T2.hadm_id
  GROUP BY T2.drug
) AS T3
WHERE T3.C1 <= 3;
\end{lstlisting}

We also include a dedicated ethical considerations section~\ref{sec:ethical} to reflect on the broader implications of lowering access barriers to clinical data via AI systems.

\section{Related Work}\subsection{Evolution of Clinical Database Access Tools}

Recent years have seen significant progress in lowering the technical barriers to accessing and analyzing complex clinical databases, particularly for researchers without advanced programming expertise. Early efforts focused on direct SQL querying and basic graphical interfaces, requiring significant technical expertise from users. The MIMIC-II project~\cite{saeed2011mimicii} introduced web-based query builders and virtual machine environments, marking an important step toward simplifying database access for clinical researchers.

The development of MIMIC-IV expanded these capabilities through various access modalities, including cloud platforms such as Google BigQuery~\cite{johnson2022mimiciv,johnson2023mimiciv}. While this improved data accessibility and processing capabilities, the fundamental challenge of SQL expertise remained and was always compounded by the equally important required understanding of the clinical domain. Visual query builders and curated SQL templates~\cite{tian2024sqlucid} have attempted to bridge this gap, though often sacrificing query flexibility for ease of use.

The emergence of standards such as HL7 FHIR, the OMOP Common Data Model, and mCODE is enabling new, more scalable methods of accessing and sharing health data. The MIMIC-IV on FHIR implementation represents an important step toward standardized data access, though it brings its own complexities in terms of resource modeling and query patterns~\cite{hl7fhir,bennett2023mimic,ohdsiStandardization}.

\subsection{Natural Language Interfaces for Medical Data}

The development of natural language interfaces for databases (NLIDB) has seen several approaches evolve in parallel. Early NLIDB implementations on healthcare domain like MIMICSQL~\cite{wang2020text} demonstrated the basic feasibility of translating natural language to SQL, though they often struggled with query complexity and medical terminology variations. Subsequent systems such as EHRSQL~\cite{yu2023ehrsql} employed more sophisticated techniques to improve query understanding, showing better handling of medical terminology while still facing challenges with complex temporal relationships and nested queries common in clinical research. A complementary line of work is retrieval-augmented generation (RAG) over biomedical corpora, benchmarked by MedRAG~\cite{xiong2024medrag}; this addresses overlapping user needs but grounds answers in retrieved documents rather than structured databases, making it a natural counterpart to the text-to-SQL approach M3 adopts.

\subsection{Benchmarks and Evaluation Frameworks}

The development of specialized benchmarks has been crucial for advancing the field. While general text-to-SQL benchmarks like BIRD~\cite{li2024can}, Spider~\cite{yu2018spider} and WikiSQL~\cite{zhong2017seq2sql} provided foundational evaluation frameworks, they lack medical domain coverage and specificity. More recent efforts such as BiomedSQL~\cite{koretsky2025biomedsql} and the EHRSQL 2024 shared task~\cite{choi2024ehrsql} have introduced domain-specific challenges that better reflect real-world clinical querying needs. Concurrent work such as TrustSQL~\cite{lee2024trustsql} has formalized the evaluation of reliability in text-to-SQL by introducing penalty-based scoring that rewards appropriate abstention on unanswerable queries, complementing the reliability framing adopted by EHRSQL 2024 and directly motivating the separate unanswerable-subset evaluation we report. These benchmarks have revealed significant challenges in handling implicit medical knowledge, understanding temporal relationships in clinical data, managing hierarchical medical concepts, and integration with clinical workflows.

\subsection{Security and Integration Frameworks}

Security considerations in clinical database access have evolved from basic database-level security and input sanitization, as outlined in resources like the OWASP SQL Injection Prevention Cheat Sheet~\cite{owasp2023sqlcheat}, to more comprehensive approaches. The introduction of the MCP~\cite{anthropic2024mcp} represents a significant advance in AI-database integration that can support modern security standards, providing precise interaction patterns, access control mechanisms, audit capabilities, and reproducible query execution. Industry adoption of MCP has indeed grown across various domains including software development, scientific research, and biomedical~\cite{mcpharma2025,mcpsuperagi2025}.

\subsection{Current Challenges and Opportunities}

Existing solutions continue to face several key challenges. General-purpose text-to-SQL systems often struggle with medical terminology and relationships, while specialized medical systems may sacrifice query flexibility for security. Many current solutions lack robust mechanisms for ensuring query provenance and result reproduction. Technical integration requirements can remain substantial, and scaling to handle the complexity of full clinical databases presents ongoing challenges. To our knowledge, none of these is currently integrated in a desktop generative AI application, such as Claude Desktop~\cite{anthropic2025desktop} for instance.

M3 builds upon these foundations while addressing these challenges through its MCP-based architecture, specialized clinical tools, and layered security design. By focusing specifically on the MIMIC-IV database and its characteristics, M3 aims to provide an accessible approach to clinical data analysis with security controls appropriate to the data.

\section{Methodology}\subsection{M3 Overview and System Architecture}

\begin{figure}[h!]
    \centering
    \includegraphics[width=0.7\linewidth]{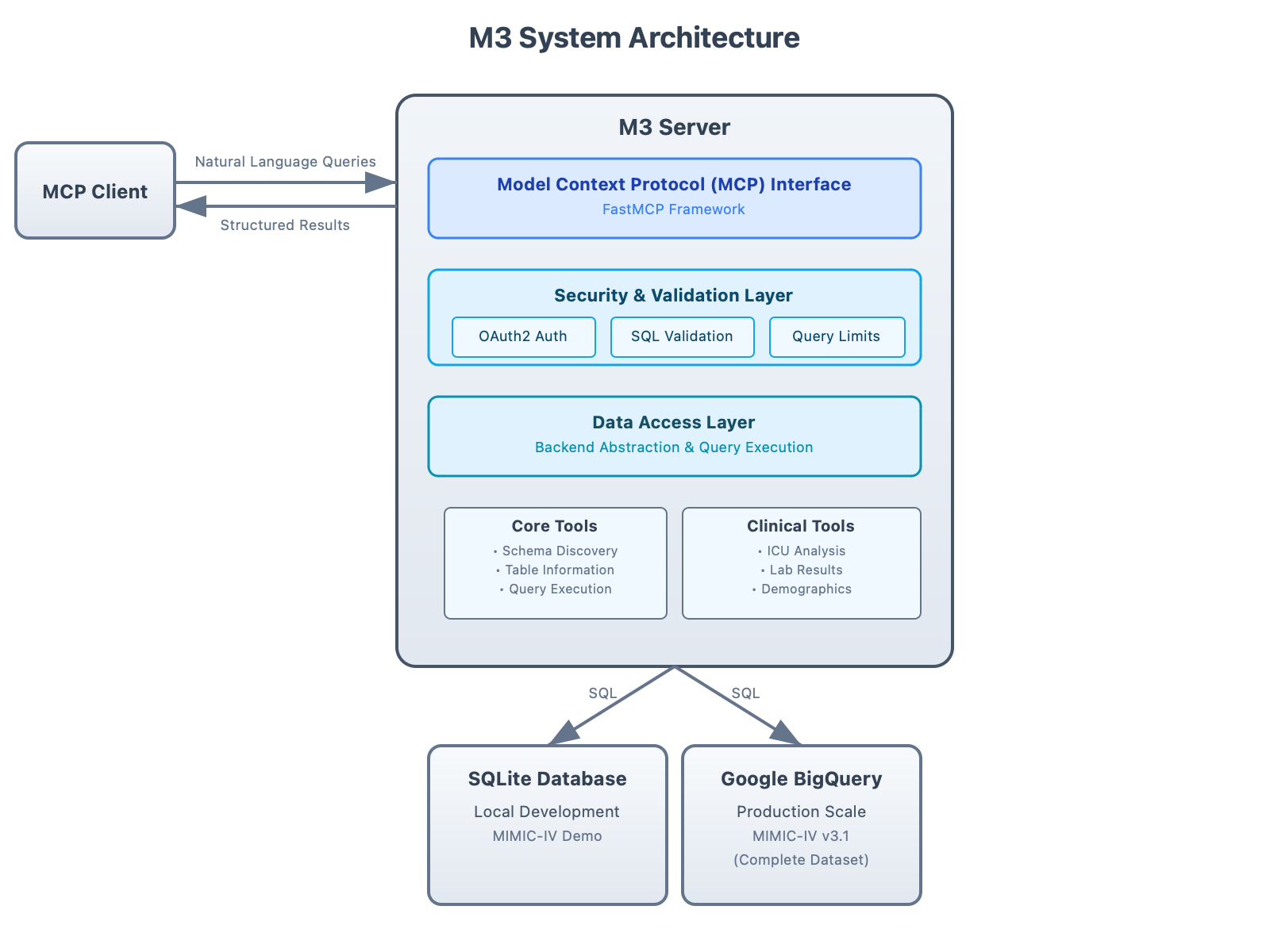}
    \caption{\textbf{Conceptual diagram of the M3 system architecture.}}
    \label{fig:m3_architecture}
\end{figure}

M3 is a Python-based server application that facilitates natural language interaction with the MIMIC-IV critical care database. Its architecture (Figure~\ref{fig:m3_architecture}) is designed around secure, modular, and user-accessible data access for clinical researchers.

The system employs a layered architecture comprising: (1) a data access layer supporting SQLite and BigQuery backends, (2) a security middleware implementing OAuth2 authentication and SQL validation, and (3) an MCP client built on the FastMCP framework that exposes tools to Large Language Model (LLM) agents. Standard software engineering best practices, such as (i) source code version control, (ii) modular architecture with abstract interfaces, (iii) functional and integration testing, are adopted across the project for ease of extension and support.

\subsection{Data Sources and Access Layer}
M3 supports two distinct backends for accessing the MIMIC-IV dataset, offering flexibility based on user needs and data scale:

\begin{itemize}
\item \textbf{Local SQLite Database}: For rapid prototyping and development, M3 provides a local SQLite implementation using the official 100-patient demo subset of MIMIC-IV~\cite{johnson2023mimiciv}. This option requires minimal setup and incurs no cloud costs. The system handles the complete Extract, Transform, Load (ETL) process from PhysioNet data files to a local database, including schema inference and standardized null value handling.

\item \textbf{Google BigQuery}: For full-scale research, M3 connects to the complete MIMIC-IV v3.1 schemas~\cite{mimicivphysionet} on Google BigQuery. This implementation supports advanced features such as parameterized queries, cost estimation, and IAM-based access control. Access requires prior PhysioNet credentialing and an active Google Cloud project.
\end{itemize}

\subsection{Configuration and Deployment}
M3 provides an interactive shell interface for system configuration and management. Users can easily select their preferred backend (SQLite or BigQuery) and configure authentication settings through this interface. The system supports both interactive and programmatic configuration approaches, allowing flexible deployment options for different research environments.

\subsection{Model Context Protocol (MCP) Tooling}
M3 exposes its functionality through a two-tiered tool architecture compatible with the Model Context Protocol. These tools enable external LLM agents to translate natural language queries into structured database operations.

\subsubsection{Core Database Tools}
The foundational layer provides essential database access capabilities including schema discovery, table inspection, and query execution. These tools allow agents to understand the database structure and execute flexible and transparent SQL queries against the MIMIC-IV dataset.

\subsubsection{Domain-Specific Clinical Tools}
To reduce complexity for common clinical research patterns, M3 provides specialized tools that encapsulate frequent operations such as retrieving ICU stay information, laboratory results, and demographic distributions. They simplify common operations while core tools provide full query transparency.

\subsection{Security and Safeguards}\label{security}
M3 is designed with a layered security framework intended to address challenges of AI-driven database access in medical research environments. The architecture includes three layers of protection. We describe these as design features; formal security evaluation (e.g., adversarial SQL-injection testing, penetration testing, and authentication audits) is identified as future work.

The authentication and authorization layer uses OAuth 2.0 with JWT tokens, enabling integration with standard identity providers. Database tools are gated by token validation; only authenticated users can invoke them.

Query validation forms the second layer through a read-only enforcement system implemented using sqlparse-based syntactic analysis. The validator blocks non-SELECT operations (including data modification or deletion) while permitting analytical queries, preserving the ability to execute legitimate research workloads.

The third layer implements resource controls. Output-size limits cap result-set sizes to prevent memory exhaustion, and per-user rate limiting with configurable thresholds manages concurrent access, supporting operational stability under expected research workloads.

\subsection{Evaluation Methodology}

We evaluated M3 using the EHRSQL 2024 benchmark~\cite{choi2024ehrsql}, a reliability-oriented text-to-SQL challenge on clinical data built on the MIMIC-IV demo database~\cite{johnson2023mimiciv_demo}. The benchmark includes both answerable questions (for which a correct SQL query and answer exist) and unanswerable questions, for which the correct system behavior is to abstain rather than produce SQL. Our primary quantitative evaluation is scoped to the answerable subset and characterizes M3's SQL generation and data retrieval accuracy. M3's behavior on the unanswerable subset is reported separately (Section~\ref{sec:unanswerable}).

\paragraph{Sampling.} The evaluation dataset was derived from the official EHRSQL 2024 MIMIC-IV test set (1,167 total questions, of which 934 are marked answerable and 233 unanswerable by the dataset's is\_answerable flag). We drew two random samples of one hundred questions each: one from the answerable subset and one from the unanswerable subset, yielding 200 evaluated questions in total. The answerable sample supports the SQL generation accuracy analysis reported below; the unanswerable sample supports the reliability evaluation in Section~\ref{sec:unanswerable}. The answerable subset was evaluated under both the Claude Sonnet 4 and gpt-oss-20B conditions, while the unanswerable subset was evaluated under the gpt-oss-20B condition to leverage the deterministic reproducibility afforded by the model's open, frozen weights. At n=100 per subset, each question receives detailed review by the multidisciplinary evaluator panel described below, and the resulting Wilson 95\% confidence intervals on the reported proportions remain sufficiently tight to support the claims drawn from them.

\paragraph{Models and deployment.} The experimental setup consisted of M3 powered by two language models. Claude Sonnet 4~\cite{anthropic2025claude4} was accessed via the Claude Desktop application at the time of evaluation (June 2025), which supports MCP servers natively. Because Claude Desktop does not expose the underlying API model revision to end users, we refer to this condition as ``Claude Sonnet 4 via Claude Desktop''. For the open-weights condition, we used LM Studio~\cite{lmstudio} to host gpt-oss-20B on a MacBook M1 Max with 32GB RAM, demonstrating feasibility of local deployment on consumer hardware. The model is \texttt{openai/gpt-oss-20b}, released by OpenAI in August 2025 under Apache 2.0 with native MXFP4 quantization~\cite{openai_gpt_oss_20b}; the exact build we ran was the LM Studio community llama.cpp GGUF~\cite{lmstudio_gpt_oss_gguf}. Because gpt-oss-20B has open weights and frozen parameters, this configuration provides deterministic reproducibility of model behavior. As with any evaluation of modern LLMs against publicly available benchmarks, the possibility of pretraining exposure to benchmark artifacts cannot be fully excluded. We provide full conversation transcripts, including tool calls and intermediate reasoning steps, so that model behavior on the specific M3 workflow can be independently inspected.

We utilized the \texttt{mimic\_iv.sqlite} database from the EHRSQL 2024 benchmark repository~\cite{ehrsql_github}, which is derived from the MIMIC-IV Clinical Database Demo v2.2~\cite{johnson2023mimiciv_demo}. The EHRSQL benchmark code defines a fixed ``current time'' of ``2100-12-31 23:59:00'' for evaluating temporal queries~\cite{ehrsql_github}. To align our M3 system with this requirement, we prepended a contextual instruction to each relevant prompt: ``Set the current time to be 2100-12-31 23:59:00 when using m3 mcp.'' This prompt-level simulation replicated the benchmark's temporal-reasoning conditions; some temporal-reasoning failures reported in our error analysis may reflect prompt sensitivity to this timestamp.

\paragraph{Scoring protocol.} The generated SQL queries and final textual answers were manually compared against the ground truth provided in the EHRSQL dataset. An answer was deemed correct if it was logically and semantically equivalent to the ground truth, even if the phrasing or presentation differed. This reliance on human judgment reflects standard practice for evaluating complex question-answering systems, as automated string-matching scripts can penalize responses that are semantically correct but differently phrased. For example, when the ground truth is ``1'' (yes), automated evaluation would incorrectly penalize a response of ``Yes, the patient meets this criterion'' despite being semantically equivalent. This approach is consistent with methodologies used in other large-scale text-to-SQL benchmarks~\cite{pourreza2023evaluating}.

\paragraph{Evaluator panel.} Correctness was adjudicated by four evaluators drawn from the research group: two with technical expertise (SQL, database schemas, and MIMIC-IV structure) and two with clinical domain expertise. Each question was reviewed independently by at least two evaluators, including at least one with technical expertise and one with clinical domain expertise, with outputs inspected in their original native formats (Claude Desktop shared conversations for Claude Sonnet 4, locally exported LM Studio JSONs for gpt-oss-20B). Disagreements were resolved by discussion among the panel, with the final per-question label reflecting consensus. No evaluator has a financial relationship with Anthropic or OpenAI.

\paragraph{Statistical analysis.} We report accuracy with Wilson 95\% confidence intervals for each model. To compare the two models on the paired set of one hundred questions, we used McNemar's test (exact binomial, two-sided) on the 2$\times$2 contingency table of per-question outcomes, the appropriate test for paired nominal data.

\paragraph{Reproducibility artifacts.} All natural language queries, model responses, generated SQL, gold-standard SQL, correctness labels, and human-annotated error notes are openly available in the M3 GitHub repository (see Section~\ref{sec:supporting}). Full conversation transcripts are provided for both conditions: Claude Sonnet 4 exchanges are linked as Claude Desktop shared conversations, and gpt-oss-20B exchanges are provided as exported LM Studio JSON transcripts.

\section{Results}\label{sec:results}

\subsection{Quantitative Performance}

On one hundred answerable questions from the EHRSQL 2024 test set, both models achieved high accuracy. Claude Sonnet 4 correctly generated SQL queries and provided the correct answer for 94 questions (94.0\%; Wilson 95\% CI [87.5\%, 97.2\%]), while gpt-oss-20B achieved 93 correct (93.0\%; Wilson 95\% CI [86.3\%, 96.6\%]).

Because both models were evaluated on the same set of questions, we compared them using a paired analysis. The resulting 2$\times$2 contingency table contains 90 questions answered correctly by both models, 4 answered correctly only by Claude Sonnet 4, 3 answered correctly only by gpt-oss-20B, and 3 answered incorrectly by both. McNemar's test on the 7 discordant pairs (exact binomial, two-sided) yields $p = 1.00$, providing no evidence of a statistically detectable performance difference between the two models at this sample size. The overlapping confidence intervals are consistent with this conclusion.

This near-equivalence is notable given that gpt-oss-20B can be deployed locally on consumer hardware, offering a viable alternative for researchers with data privacy requirements or limited cloud connectivity. We emphasize that these results are scoped to the answerable subset of the benchmark; M3's behavior on unanswerable questions --- where correct system behavior is to abstain rather than produce SQL --- is evaluated separately (see Section~\ref{sec:unanswerable}).

\begin{table}[h!]
\centering
\caption{Evaluation Results on a 100-Sample Subset of the EHRSQL Test Set (answerable questions).}
\label{tab:results}
\begin{tabular}{lcc}
\hline
\textbf{Outcome} & \textbf{Claude Sonnet 4} & \textbf{gpt-oss-20B} \\ \hline
Correct Answers & 94 & 93 \\
Incorrect Answers & 6 & 7 \\ \hline
\textbf{Total Evaluated} & \textbf{100} & \textbf{100} \\ \hline
\end{tabular}
\end{table}

\subsection{Model Comparison and Error Analysis}

Analysis of the incorrect responses reveals both shared failure modes and model-specific weaknesses. Understanding these patterns provides insight into the limitations of current natural language-to-SQL systems in clinical contexts.

\textbf{Common failures across both models.} Three questions challenged both Claude Sonnet 4 and gpt-oss-20B, suggesting these questions contain inherent ambiguities or make unstated assumptions.

One question asked about days elapsed since a patient's last discharge lounge stay. The gold query calculated from entry time (intime), while both models used exit time (outtime). The phrase ``days since last stay'' does not explicitly specify whether the reference point is the beginning or end of the stay, making both interpretations defensible.

Another question about the most frequent microbiology tests for hemodialysis patients did not specify the temporal relationship between tests and procedure. The gold query required tests to occur after hemodialysis during the same admission, but this ordering was not explicit in the question wording. Both models returned different test lists, with some overlap. The discrepancy appears to stem from how each model handled cases where multiple tests had identical frequencies, leading to different selections among equally-ranked options.

A third question asked about patients receiving serology/blood microbiology tests since 2100. Both models overcounted substantially (Claude: 41 patients, gpt-oss-20B: 40 patients) versus the gold standard (8 patients). The gold query searched for the exact specimen type `serology/blood', while both models interpreted this phrase more broadly to include various blood-related tests. Without explicit enumeration of valid specimen type values, this broader interpretation represents a reasonable approach to an underspecified query.

\textbf{Claude Sonnet 4-specific errors.} Claude made three unique errors. A question asking for the difference between the last and second-to-last blood pressure measurements was incorrectly interpreted, with Claude comparing the first and second-to-last values instead, yielding +8.0 mmHg instead of the correct -9 mmHg. This appears to be a temporal ordering mistake in query construction.

A question about drugs prescribed after alcohol detoxification during the same hospital visit resulted in Claude identifying only phenobarbital-related medications while omitting other concurrent prescriptions (docusate sodium, haloperidol, latanoprost, omeprazole, sarna lotion). The model appears to have filtered for withdrawal-related drugs, though the question did not specify this restriction.

For a question asking for the count of current patients aged 60 or above, Claude returned 44 patients by counting those without a death date, while the correct interpretation required checking for ongoing hospitalization (dischtime IS NULL), which would yield 1 patient. This reflects the need for more precise terminology, distinguishing ``currently admitted'' from ``alive'' (not deceased) to avoid ambiguity in database queries.

\textbf{gpt-oss-20B-specific errors.} gpt-oss-20B made four unique errors. For a question about diagnoses following BMI 35.0-35.9 diagnosis, the model used ICD-9 code 278.00 (general obesity) as a proxy, explaining that BMI data was not directly available. However, the database actually contained a specific ICD code for ``body mass index 35.0-35.9, adult'' that the model failed to discover through schema exploration.

A question about patients receiving nutritional substance introduction after postprocedural pneumothorax resulted in gpt-oss-20B returning 0 patients while the ground truth indicated 1. This likely resulted from using hardcoded ICD codes rather than the gold query's flexible approach of matching long title descriptions.

For a question asking for the first specimen test given to a patient since March 2100, gpt-oss-20B incorrectly identified ``pt'' (prothrombin time) from the labevents table, while the correct answer ``mrsa screen'' resided in the microbiologyevents table. This indicates confusion between laboratory test results and specimen collection procedures, which are tracked in different tables.

Finally, for a question asking for the top four most frequent lab tests, gpt-oss-20B's results (glucose, chloride, sodium, hemoglobin) differed from the gold standard (chloride, creatinine, hematocrit, sodium). The discrepancy likely stems from different approaches to handling frequency ties, as the gold query uses DENSE\_RANK which can return a different number of results than expected when multiple tests have the same frequency.

\textbf{Error taxonomy.} Table~\ref{tab:error_taxonomy} summarizes the distribution of errors across both models. Question ambiguity accounted for the majority of errors, including all three cases in which both models failed. Schema-exploration gaps and frequency tie-breaking discrepancies were specific to the gpt-oss-20B condition, while the single temporal-logic error was specific to Claude Sonnet 4.

\begin{table}[h!]
\centering
\caption{\textbf{Error taxonomy by category and model.} Counts reflect the ten questions from the 100-question evaluation on which at least one model produced an incorrect answer. The ``Shared'' column counts questions on which both models failed; ``Claude only'' and ``gpt-oss-20B only'' count questions on which only that model failed.}
\label{tab:error_taxonomy}
\resizebox{\textwidth}{!}{%
\begin{tabular}{lcccc}
\hline
\textbf{Category} & \textbf{Shared} & \textbf{Claude only} & \textbf{gpt-oss-20B only} & \textbf{Total}\\
\hline
Question ambiguity / underspecification & 3 & 2 & 0 & 5\\
Schema exploration gap & 0 & 0 & 3 & 3\\
Temporal logic & 0 & 1 & 0 & 1\\
Frequency tie-breaking (\texttt{DENSE\_RANK}) & 0 & 0 & 1 & 1\\
\hline
\textbf{Total} & \textbf{3} & \textbf{3} & \textbf{4} & \textbf{10}\\
\hline
\end{tabular}%
}
\end{table}

(See Section~\ref{sec:supporting} for access to complete benchmark results and error analysis.)

\subsection{Reliability Evaluation on Unanswerable Questions}
\label{sec:unanswerable}

The EHRSQL 2024 benchmark is designed as a reliability task: the test set contains unanswerable questions for which the correct system behavior is to abstain rather than produce SQL. We evaluated M3's behavior on this axis using the gpt-oss-20B condition, which provides deterministic reproducibility through frozen open weights and enables exact replication by other researchers without dependence on subscription-based model access.

Each of the 100 sampled unanswerable questions was submitted to M3 and the resulting response was adjudicated by the evaluator panel into one of three outcomes: correctly abstained, SQL-backed answer that misrepresents the data, or general-knowledge answer that does not abstain. The model correctly abstained on 69 of 100 questions (69.0\%; Wilson 95\% CI [59.4\%, 77.2\%]). Table~\ref{tab:unanswerable_patterns} summarizes the outcome distribution.

\begin{table}[h!]
\centering
\caption{\textbf{Outcome distribution on the unanswerable subset (gpt-oss-20B, n=100).} For unanswerable questions, the correct behavior is abstention; the remaining categories represent distinct reliability failure modes.}
\label{tab:unanswerable_patterns}
\begin{tabular}{lc}
\hline
\textbf{Outcome} & \textbf{Count}\\
\hline
Correctly abstained & 69\\
SQL-backed answer with mismatched schema mapping & 12\\
General-knowledge response without abstention & 15\\
Other (clarification requests, misinterpretations) & 4\\
\hline
\textbf{Total} & \textbf{100}\\
\hline
\end{tabular}
\end{table}

The 12 SQL-backed failures share a common pattern. The model runs a valid SQL query against real data but labels the result under a name that does not match the question's intent. For example, asked for ``the invoice numbers of the female patients who are not deceased'', the model joins the \texttt{cost} and \texttt{patients} tables and returns distinct \texttt{event\_id} values aliased as invoice numbers, although \texttt{event\_id} is not an invoice identifier. Similar mismatches appear in questions about locations among hospitals (returned as \texttt{admission\_location} categories in a single-hospital dataset) and about the ``short title'' of an ICD-9 code (returned as the leading words of the real \texttt{long\_title} column). This is the same underlying failure surfaced by the answerable-subset error taxonomy (Table~\ref{tab:error_taxonomy}): the model commits to a reasonable-sounding mapping between a natural-language entity and a schema element that does not semantically match.

The 15 general-knowledge responses are typically clinical or pharmacological explanations produced in reply to prompts that resemble information requests but are not grounded in the MIMIC-IV schema (for instance, summaries of propofol side effects or treatment recommendations for atrial fibrillation). Such answers may be factually reasonable in isolation but do not constitute correct behavior under the reliability-oriented framing of the benchmark, which treats any non-abstention on an unanswerable question as an error.

The 69\% abstention rate establishes a baseline reliability profile for M3 paired with a small open-weights model, and identifies the dominant correctable failure mode as the commitment to mismatched schema mappings. Pairing M3 with an explicit ambiguity-detection or clarifying-question step before SQL generation is a concrete direction for reducing reliability failures. Full conversation transcripts for all 100 items, per-question classifications, and reviewer notes are available in the M3 GitHub repository (see Section~\ref{sec:supporting}).

\subsection{Visual Examples}

\begin{figure}[h]
  \centering
  \begin{minipage}[c]{0.48\linewidth}
    \centering
    \includegraphics[width=\linewidth]{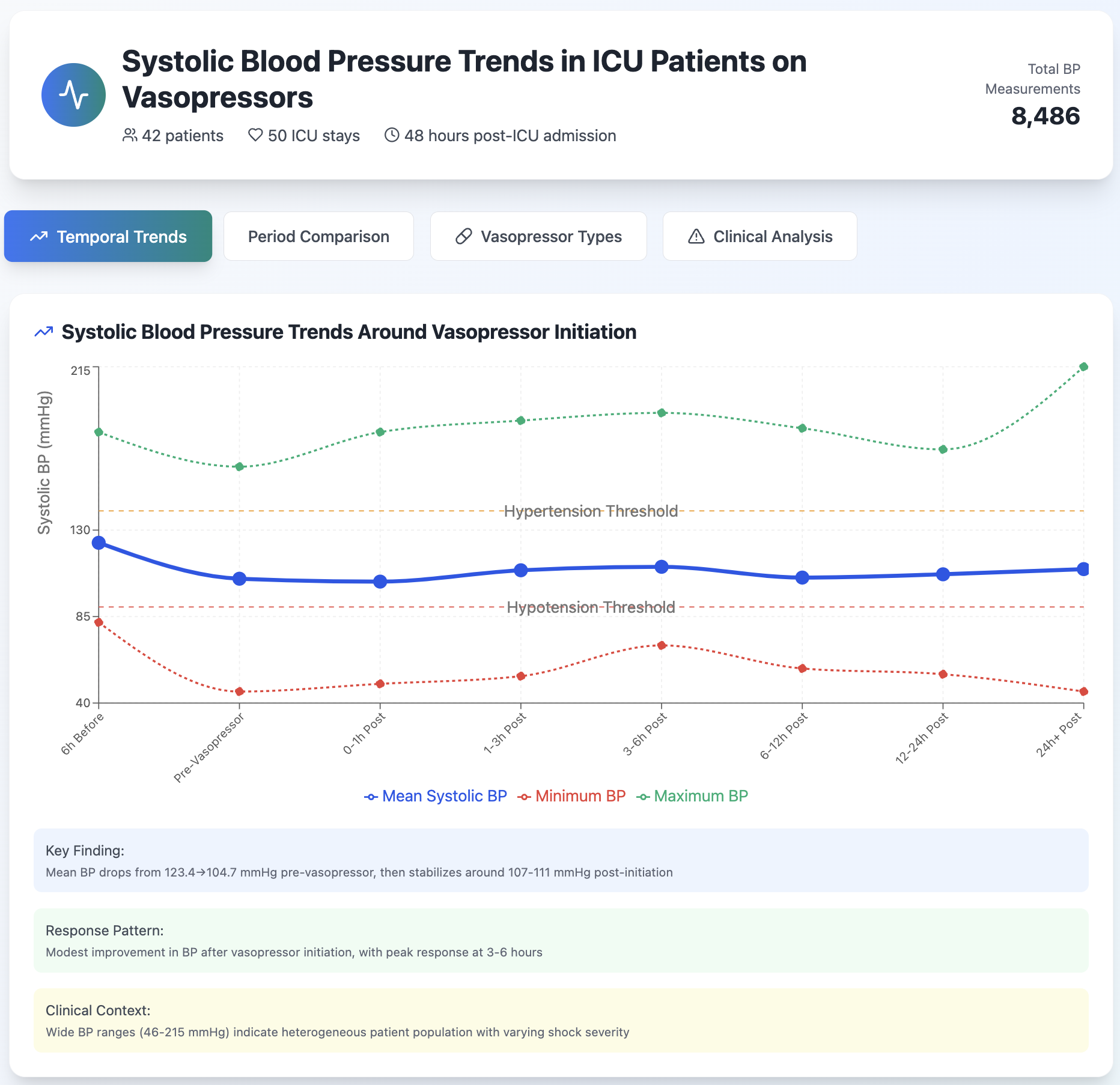}
    \captionof{figure}{\textbf{Systolic blood pressure trends for patients on vasopressors.} Query: \textit{``Show trends in systolic blood pressure for patients on vasopressors within 48 hours of ICU admission.''}}
    \label{fig:query1}
  \end{minipage}
  \hfill
  \begin{minipage}[c]{0.48\linewidth}
    \centering
    \includegraphics[width=\linewidth]{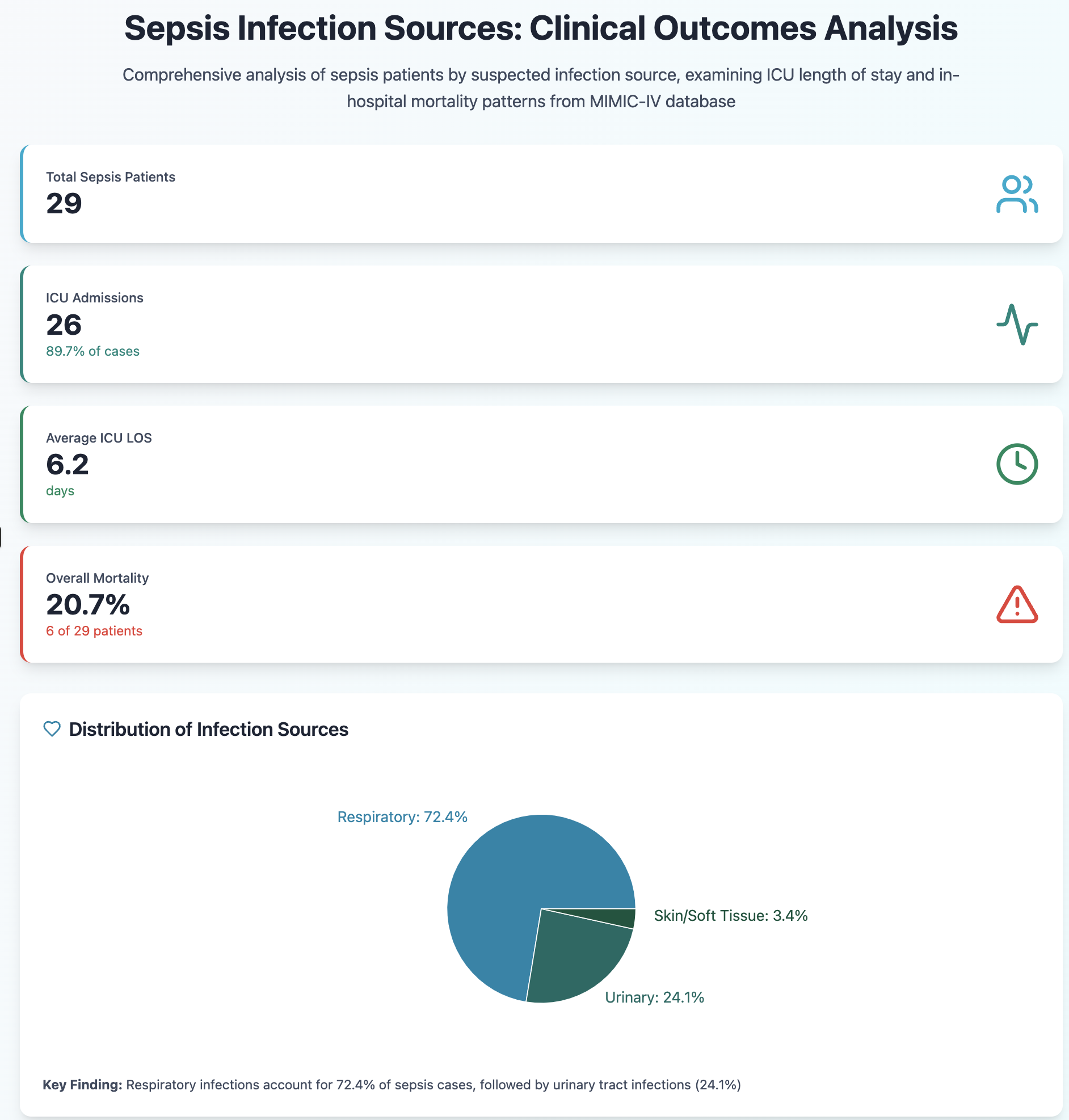}
    \captionof{figure}{\textbf{Sepsis patient analysis by infection source.} Query: \textit{``Among sepsis patients, what's the source-of-infection distribution and how do groups differ in ICU stay and mortality?''}}
    \label{fig:query2}
  \end{minipage}
\end{figure}

To complement the quantitative evaluation, Figures~\ref{fig:query1} and~\ref{fig:query2} present illustrative examples of complex queries processed by M3 on the MIMIC-IV demo~\cite{johnson2023mimiciv_demo}. These were generated using the MIMIC-IV demo database via the Claude-powered M3 system. Each example includes the natural language query and the resulting visualization, designed to reflect real-world clinical insights extractable from MIMIC-IV.

These outputs highlight M3's ability to retrieve data and present it in formats immediately interpretable to clinicians and researchers.

\subsection{Discussion of Results}

The evaluation results offer encouraging preliminary validation of the M3 architecture. The 93-94\% accuracy demonstrates that language models, when provided with proper tools via MCP, can effectively query complex databases like MIMIC-IV without task-specific fine-tuning.

The near-equivalent performance of gpt-oss-20B has important implications. For researchers facing data privacy constraints, regulatory requirements, or limited connectivity, local deployment offers a practical path forward. The smaller parameter count contributes to lower computational requirements and faster inference on local hardware.

The qualitative analysis underscores that the primary challenge is semantic: correctly interpreting user intent. Errors caused by linguistic ambiguity suggest future work should focus on ambiguity detection and resolution~\cite{min2020ambigqa}. The fact that both models encountered similar obstacles suggests these represent general challenges for natural language-to-SQL in clinical contexts~\cite{singhal2023large}.

Most errors could potentially be resolved through multi-turn conversations where user and system iteratively refine query specification. For basic exploratory data analysis, M3 performs remarkably well. For complex analytical tasks involving nuanced temporal relationships, expert involvement remains valuable.

The transparency of M3's approach, exposing generated SQL alongside natural language results, enables expert oversight. Researchers can verify that interpretations match intent and spot logical errors, maintaining analytical rigor.

\section{Ethical Considerations}\label{sec:ethical}

The development and deployment of AI systems like M3 occur within societies characterized by profound forms of social, material, and political inequality. The healthcare and medical research domains are particularly susceptible to these inequalities, making it essential to address the ethical implications of technologies that democratize access to clinical data analysis~\cite{birhane2021algorithmic}.

\subsection{Benefits and Maintaining Analytical Rigor}

M3 offers significant potential benefits by making clinical data more accessible to researchers, including those with limited computational resources or laboratory infrastructure for complex data analyses. This accessibility could advance medical knowledge and help address system-level inequalities by enabling broader participation in clinical research. The democratization of clinical data analysis represents an opportunity to engage diverse perspectives in medical research while maintaining appropriate safeguards.

The extensive training and experience previously required to conduct database queries traditionally enabled research scientists and clinicians to evaluate the scientific validity of their analyses, identify potential misinterpretations of statistical results, and understand the complexities of translating query results into clinical practice~\cite{singhal2023large}. Experienced researchers typically possess intimate knowledge of the datasets they work with, including understanding how specific categories were constructed, how data was collected, and the implications of these factors for specific research queries.

To ensure M3 users can maintain this level of analytical rigor, we recommend implementing comprehensive training programs that bridge the gap between technical accessibility and domain expertise. This includes providing detailed documentation about dataset construction, establishing mentorship programs pairing experienced researchers with new users, and creating educational resources that emphasize the importance of contextual understanding in clinical data analysis.

M3's design principle of exposing underlying SQL queries alongside natural language results directly supports this goal by enabling users to understand and validate the analytical approach, fostering transparency and reproducibility in research workflows.

\subsection{Promoting Equity and Addressing Bias}

The past decade has witnessed extensive focus on bias in artificial intelligence systems, with AI algorithms shown to replicate and amplify existing forms of societal inequality and discrimination~\cite{birhane2021algorithmic}. These concerns are particularly acute in healthcare, where biased algorithms can perpetuate historical injustices and disproportionately affect communities already experiencing significant social and health inequalities~\cite{obermeyer2019dissecting}.

M3 users should be equipped with training and tools to identify how specific populations, particularly marginalized groups, may be represented in datasets, and how to conduct analyses that account for potential biases. To support this, we recommend developing bias awareness training, implementing tools that help users understand the demographic composition of their analyses, and establishing review processes that evaluate research for potential equity implications.

The transparency provided by M3's query exposition enables peer review and validation of analytical approaches, supporting the identification and correction of potential biases in research design and interpretation.

\subsection{Security, Privacy, and Accountability}

M3 incorporates several design-level safeguards aimed at privacy and security concerns inherent in clinical data analysis. While MIMIC-IV consists of de-identified data, M3's architecture includes OAuth2 authentication, read-only query validation intended to prevent unauthorized operations, audit logging, and per-user rate limiting. These are design features rather than empirically validated security properties; formal security evaluation remains future work.

The availability of local deployment via open-weights models like gpt-oss-20B offers an additional privacy advantage. Institutions with strict data governance requirements can run M3 entirely within their own infrastructure, ensuring that query patterns, intermediate results, and analytical approaches never leave their controlled environment. This local deployment option addresses concerns about inadvertent information leakage through cloud-based API calls while maintaining the benefits of natural language interaction.

To ensure accountability in AI-assisted research, we propose a collaborative responsibility model where system developers maintain robust security measures and clear documentation, users employ the system appropriately with proper training, institutions establish governance frameworks, and the research community maintains quality standards through peer review.

The linguistic ambiguity challenges identified in our evaluation results (Section~\ref{sec:results}) highlight the importance of verification procedures. M3's query transparency features enable users to validate their analyses and support reproducible research practices. We recommend that institutions establish procedures for reviewing AI-generated analyses, particularly those intended for clinical application or publication.

\subsection{Implementation and Best Practices}

To ensure M3 enhances rather than compromises scientific rigor, we recommend several best practices. Users should validate results through multiple approaches where possible, thoroughly document their analytical procedures including the natural language queries used, and ensure appropriate peer review of their work. M3's transparency features, including exposed SQL queries and comprehensive logging, directly support these practices.

Training programs should emphasize the importance of critical evaluation skills and help users understand both the capabilities and limitations of AI-assisted analysis. By combining M3's accessibility with robust educational frameworks, we can democratize clinical data analysis while maintaining the highest standards of scientific excellence.

Based on these considerations, we recommend institutions adopting M3 implement phased deployment strategies, beginning with supervised use in educational settings. Comprehensive training programs should address both technical and ethical aspects of AI-assisted clinical data analysis. Clear governance frameworks should establish policies for M3 usage, including guidelines for result interpretation and approval processes for sensitive analyses.

Regular monitoring of M3 usage patterns and outcomes can help identify areas for improvement and ensure alignment with institutional and professional standards. Engaging diverse stakeholders, including M3 users, clinical experts, and ethicists, will help ensure ongoing alignment with evolving best practices.

M3 represents a significant opportunity to democratize clinical data analysis while maintaining the rigor essential for advancing medical knowledge. Through careful attention to ethical considerations, comprehensive training, and robust governance frameworks, we can harness the benefits of this technology while preserving the integrity of medical research and promoting equitable access to clinical insights.
\section{Conclusion and Future Work}\subsection{Conclusion}

M3 demonstrates that a protocol-driven natural language interface to complex clinical databases is feasible and can reduce friction in exploratory research workflows. The system's dual-backend architecture supports local SQLite databases for rapid prototyping and connects to Google BigQuery for full-scale deployment; empirical evaluation on the full MIMIC-IV dataset is identified as future work. M3 exposes two tiers of MCP tools, combining foundational database operations (schema discovery, query execution) with domain-specific clinical functions, with the goal of bridging raw SQL capabilities and medical research workflows. The dual-backend design also serves an educational function, allowing students and researchers to learn clinical data analysis techniques on a local demo dataset before scaling.

The system aims to lower the technical barrier for researchers who lack SQL expertise or schema-level familiarity. M3 is designed with security and reproducibility features suited to sensitive medical data: layered query validation, OAuth2-based access control, and rate limiting. These controls, implemented through sqlparse-based validation and JWT token authentication, are aligned with OWASP recommendations as design references; their effectiveness under adversarial conditions has not been formally evaluated and represents future work.

At the same time, we acknowledge that M3 is only a starting point. Its current focus on MIMIC-IV, dependence on LLM quality, and narrow focus on data retrieval highlight opportunities for deeper integration with broader research and clinical workflows. Nonetheless, the successful deployment of M3 affirms that such interfaces can meaningfully reduce friction in data exploration, and we hope this work inspires continued development and community-driven extension.

\subsection{Limitations}
\label{sec:limitations}

Several limitations warrant attention. First, empirical evaluation was conducted on the 100-patient MIMIC-IV demo subset; evaluation on the full MIMIC-IV v3.1 dataset via the BigQuery backend is an important next step, as scale-dependent differences in query complexity, schema coverage, and runtime behavior remain to be characterized at full dataset size.

Second, SQL queries that execute successfully but reflect an incorrect interpretation of the question can be difficult for non-expert users to detect, since the returned results appear well-formed. MCP-compatible clients typically display the generated SQL and tool-call trace alongside the natural-language answer, which makes expert review possible; user-facing deployments should nonetheless pair the system with appropriate oversight rather than treat results as authoritative in isolation.

Third, while the benchmark evaluation characterizes SQL generation accuracy on standardized questions, end-user studies with practicing clinical researchers in real analytical workflows are left to future work. Such studies are needed to assess how the interface performs in everyday research practice and how users interpret generated queries and results.

Finally, institutional deployments will need to adapt authentication, data governance, and audit protocols to local policies; the current OAuth2/JWT configuration is a reference implementation rather than a prescription, and richer identity-provider integration is included on the roadmap.

\subsection{Roadmap}

We invite the research community to participate in the development of M3, submitting Pull Requests on the official GitHub repository: \url{https://github.com/rafiattrach/m3}. Here are the list of priorities, as identified by M3 stakeholders, where we welcome immediate contributions:

\paragraph{A. Broader Dataset Coverage.} One of our immediate priorities is expanding M3 beyond MIMIC-IV. Planned connectors include additional PhysioNet datasets (e.g., MIMIC-CXR, MIMIC-IV-ED), multi-institutional tabular repositories like eICU, and FHIR-compatible formats. This will require a modular ingestion layer capable of abstracting over heterogeneous schemas while exposing a unified natural language interface. This expansion will be accompanied by performance optimizations including query result caching, connection pooling, and intelligent query routing to minimize latency and computational costs across diverse backend systems.

\paragraph{B. Richer MCP Tooling.} Future M3 versions will extend the MCP interface to include not only core SQL capabilities but also higher-level clinical tasks. These include cohort definition tools, summarization functions, declarative visualization endpoints, and retrieval-augmented generation (RAG) utilities for grounding responses in biomedical literature. Each of these will be exposed as an explicit MCP tool with well-scoped permissions.

\paragraph{C. Technical Enhancements.} Several technical improvements will strengthen M3's robustness and performance. Advanced rate limiting with adaptive thresholds based on query complexity will optimize resource utilization beyond the current per-user request counting approach. Query result caching and connection pooling will improve response times for frequently accessed data patterns. Additionally, expanded authentication provider support beyond the current OAuth2/JWT implementation will accommodate diverse institutional identity management systems.

\paragraph{D. Ecosystem and Community Contributions.} We envision M3 evolving into a community platform for natural language-driven clinical research. To support this, we plan to introduce a plugin system and formalize contribution guidelines, including continuous integration pipelines to validate third-party ingestion, query, or analysis modules against test datasets.

Together, these enhancements will move M3 from a research prototype toward a robust, extensible foundation for secure, language-driven interaction with clinical data systems.

\FloatBarrier   

\section*{Competing interests}

The Model Context Protocol (MCP) used by M3 is an open protocol developed by Anthropic. One of the two models evaluated, Claude Sonnet 4, is also developed by Anthropic; the other, gpt-oss-20B, is an open-weights model released by OpenAI. The authors have no financial relationship with either Anthropic or OpenAI and are not employed or compensated by any AI-model provider.

\section*{Acknowledgments}

The authors thank Dr. Gloria Hyunjung Kwak for clinical domain expertise. PM acknowledges financial support from the Fulbright Scholarship and Erasmus Mundus JM Scholarship. LAC is funded by the National Institute of Health through DS-I Africa U54 TW012043-01 and Bridge2AI OT2OD032701, the National Science Foundation through ITEST \#2148451, and a grant of the Korea Health Technology R\&D Project through the Korea Health Industry Development Institute (KHIDI), funded by the Ministry of Health \& Welfare, Republic of Korea (grant number: RS-2024-00403047). This research was supported by a grant of the Korea Health Technology R\&D Project through the Korea Health Industry Development Institute (KHIDI), funded by the Ministry of Health \& Welfare, Republic of Korea (grant number: RS-2024-00439677).

M3 operates exclusively on de-identified MIMIC-IV released through PhysioNet. All investigators completed required CITI training. All empirical results were obtained from publicly available resources.

\bibliographystyle{unsrtnat}
\bibliography{m3}

\clearpage
\appendix
\newpage
\section*{CRediT Author Statement}

Rafi Al Attrach: Investigation (lead), Software (lead), Writing – original draft (lead), Writing – review \& editing (equal).

Pedro Moreira: Investigation (lead), Software (lead), Writing – original draft (lead), Funding acquisition (supporting), Writing – review \& editing (equal).

Rajna Fani: Investigation (lead), Software (lead), Writing – original draft (lead), Writing – review \& editing (equal).

Renato Umeton: Conceptualization (equal), Methodology (supporting), Investigation (supporting), Writing – review \& editing (equal).

Amelia Fiske: Conceptualization (supporting), Writing – review \& editing (equal), Ethics (lead).

Leo A. Celi: Conceptualization (equal), Supervision (lead), Project administration (lead), Writing – review \& editing (equal).

This statement is based on CRediT, the \href{https://credit.niso.org/}{ANSI/NISO Contributor Role Taxonomy}.

\section*{Supporting Information}
\label{sec:supporting}

Detailed benchmark results including all natural language queries, model answers, gold standard answers, SQL queries, and error analysis notes are publicly available in the M3 GitHub repository:

Claude Sonnet 4 answerable-subset results: \url{https://github.com/rafiattrach/m3/blob/main/benchmarks/ehrsql-naacl2024/claude-sonnet-4/EHRSQL_benchmark.csv}

gpt-oss-20B answerable-subset results: \url{https://github.com/rafiattrach/m3/blob/main/benchmarks/ehrsql-naacl2024/gpt-oss-20B/EHRSQL_benchmark.csv}

gpt-oss-20B unanswerable-subset results: \url{https://github.com/rafiattrach/m3/blob/main/benchmarks/ehrsql-naacl2024/gpt-oss-20B/EHRSQL_benchmark_unanswerable.csv}

\end{document}